\documentclass{article}
\usepackage{graphicx} 
\usepackage{hyperref}
\usepackage{floatrow}
\usepackage{adjustbox}

\hypersetup{
    colorlinks=true,
    linkcolor=black,
    urlcolor=blue
}

\title{Extending Interactive Science Exhibits into the Classroom using Anthropomorphized Chatbots and Bloom’s Taxonomy}
\author{Yousuf Golding\thanks{UC Santa Cruz}}
\date{30th January 2024}

\begin{document}

\maketitle

\section*{Abstract}

\paragraph{This study explores the use of Generative AI chatbots for transforming public science exhibits into virtual experiences that can extend the engagement of exhibits into the classroom. The broader goal is to increase accessibility of science exhibits, especially for those marginalized in STEM due to various factors, including cultural barriers. We hypothesize that turning exhibits into first-person anthropomorphized chatbots with a personality, like quirky-talking asteroids or comets, can increase engagement and learning. The paper mainly explores if such techniques are possible using Generative AI (e.g. GPT) via prompt engineering alone. The research includes an investigation into the possibility of integrating interactive assessment via question-generation using Bloom’s Taxonomy. Initial results indicate that it is possible to combine these techniques. As such, it lays a foundation for future classroom evaluations of such chatbots to gauge their overall efficacy in extending the reach of science exhibitions. The paper concludes by discussing extensions of the research to fully evaluate effectiveness in virtual  field-trips. We also include a brief examination of additional ways to enhance student motivation towards learning via chatbots.}

\section*{Introduction}

\paragraph{High school and middle school students benefit from learning about science via hands-on exhibits and demonstrations in public-science establishments such as museums and centers like the \href{https://chabotspace.org/}{Chabot Space and Science Center website} (CSSC), Oakland. Field trips enable students to develop interest in science, which may lead to improved learning or improved science literacy (Behrendt et al., 2014) and improvements in science-related test scores (Whitesell, 2016)}

\paragraph{An obvious question is how to extend the reach of such exhibits into the classroom using digital technology. By providing a more cost-effective alternative to in-person visits, virtual exhibits could enhance the reach of such centers (Behrendt et al., 2014). Ideally, given the value of hands-on or participatory demonstrations (Ekwueme et al., 2015), virtual exhibits would attempt to incorporate features that could encourage engagement with science. Also, AI tools have the potential to improve student success and engagement, particularly among those from disadvantaged backgrounds (Sullivan et al., 2023) and those marginalized from STEM due to cultural exclusion (Cook, 2023).}

\paragraph{To that end, our hypothesis is that AI-powered chatbots, if designed correctly, might provide a suitable basis for “virtual exhibits”, given their interactive nature. However, interactivity in of itself isn’t guaranteed to be engaging. We therefore explored potential mechanisms for elevating engagement. Research revealed two potential mechanisms: persona-based chat (Dwivedi et al., 2023) and interactive assessment via questions that exploit Bloom’s taxonomy of learning (Adams, N. 2015)}

\paragraph{Our longer-term research question, if we had the time and resources, would be to evaluate how effective the proposed chatbot-related ideas are in engaging students and enhancing learning outcomes in the field. However, this would require a full field test with the participation of students and their teachers, and possibly the involvement of a science center like the CSSC. }

\paragraph{Given our limited resources, we chose instead to evaluate the feasibility of using Generative AI (GenAI) chatbots to achieve the following:}

\begin{enumerate}
    \item Ease of production so that educationalists, including voluntary ones in science centers, could easily produce a chatbot without knowing programming.
    \item Feasibility of incorporating anthropomorphic features, in particular:
    \begin{itemize}
        \item Can GenAI allow the exhibit to behave in the first-person as if the user is chatting to the exhibit?
        \item Can GenAI enable the exhibit to take on a particular persona in order to inject a personality?
    \end{itemize}
    \item Feasibility of incorporating assessment via question generation (QG) using Bloom’s Taxonomy–can GenAI generate suitable questions automatically?
\end{enumerate}

\section*{Background}

\subsection*{Chatbots to Enhance Learning}

\paragraph{The use of chatbots in education is not new. Literature surveys suggest broad applicability to a range of educational circumstances (Pérez et al., 2020). Additionally, impacts upon some educational outcomes have been shown to be positive (Rong et al., 2023) such as the ability to provide per-student personalized learning (Winkler et al., 2018). According to Vazquez-Cano et al. (2021), a well-designed chatbot can make learning more continuous and automatic. The use of chatbots within the context of “micro-learning” has shown them to be effective in enhancing motivations towards learning (Yin et al., 2021). In a study to explore learning English vocabulary (as a foreign language) results showed that vocabulary gains in the test group were significantly higher than in control groups (Annamalai et al, 2023).}

\subsection*{Anthropomorphic Design}

\paragraph{Chatbots can be improved in terms of educational engagement by equipping them with human-like features by incorporating anthropomorphic design features (Dwivedi et al., 2023). Anthropomorphism can take the form of embodied designs, such as human 3D representations, or disembodied interfaces, such as text-only interfaces, but nonetheless still imbued with human traits, like “personality”. Many forms of anthropomorphism for chatbot designs have been attempted (Janson et al., 2023).}

\paragraph{These stylistic modifications can include personification (Pizzi, Scarpi, \& Pantano, 2021) to create what is sometimes called personality-adaptive chatbots (Ait Baha, Tarek, et al 2023). Due to recent technological developments such as generative AI (Bommasani et al., 2021), adapting chatbots to include stylistic influences is possible via creative use of inputs (prompts), such as “write me poem in the style of grime rap”. This lays the foundations for enabling personified bot dialogs.}

\subsection*{Learning Assessment}

\paragraph{Research by Annamalai et al., (2023) revealed that chatbots in the classroom supported competence, autonomy, and relatedness, which are aspects of Self Determination Theory, a psychological framework for developing motivation. In that study, students placed value on the use of chatbots in helping with assessment within a blended-learning approach. In terms of assessment, the development of meta-cognition has been shown to enhance learning outcomes via the use of questions derived from Bloom’s Taxonomy (BT) (Sudirtha et al., 2022)}

\paragraph{BT includes six categories in the revised edition:}

\begin{itemize}
    \item \textbf{Remembering}: Retrieving, recognizing, and recalling relevant knowledge from long-term memory.
    \item \textbf{Understanding}: Constructing meaning from oral, written, and graphic messages.
    \item \textbf{Applying}: Carrying out or using a procedure through executing or implementing.
    \item \textbf{Analyzing}: Breaking material into constituent parts and detecting how the part relate to one another and to an overall structure or purpose.
    \item \textbf{Evaluating}: Making judgments based on criteria and standards.
    \item \textbf{Creating}: Putting elements together to form a coherent or functional whole; reorganizing elements into a new pattern or structure.

\end{itemize}

\paragraph{We performed experiments in QG initially with all six categories, but constrained the hybrid testing of QG with personified chatbots to only the first two as this was easier to evaluate and to control during AI generation. Also, for reasons that will become clear, the use of higher-order categories introduce certain complexities in the way the final solution might need to be designed and delivered into the classroom. Note that we did not explore the many criticisms of BT (Kompa, 2017), except for one of them: that learning is non-linear and doesn’t necessarily follow the neat ascent of the taxonomy. This criticism was of interest because of the capabilities of modern GenAI chatbot techniques in accommodating non-linear learning paths (see later).}

\subsection*{Chabots using Generative AI and ChatGPT}

\paragraph{AI has recently undergone a transformation due to the invention of Generative AI (GenAI) (Garrido-Merchán et al., 2023). Our study utilized the latest GenAI technology (GPT-3.5) that powers ChatGPT. This technology can capture and retain contextual information throughout interactions, leading to more student-relevant conversations. Unlike previous-generation chatbots that follow fixed learning paths (or decision trees), ChatGPT can engage in open-ended dialogue. This seems more compatible with our goals to provide a more engaging experience as it allows the student some degree of autonomy.}

\paragraph{Moreover, its adaptability allows it to accommodate different language styles, and even write and debug computer code, making it a valuable tool in educational settings (Baidoo-Anu \& Owusu Ansah, 2023; Tate et al., 2023). One such adaptability is the ability to incorporate stylistic influences, such as the use of personas to add a particular voice (“personality”) to the chat, thus aiding our goal to incorporate anthropomorphic features into the user experience.}

\section*{Methods}

\paragraph{The tests in this work were conducted by prompting \textit{GPT-3.5-turb}o and \textit{GPT-3.5-turbo-instruct}. Both were used because of our use of the legacy mode in the OpenAI playground, which supports the former. We also used ChatGPT (3.5) to generate example text passages to compare with passages scraped from the NASA educational guide (e.g. \href{https://www.jpl.nasa.gov/edu/teach/activity/modeling-an-asteroid/}{Modeling an Asteroid activity}). Due to limited time, we only explored a single subject, namely the topic of comets and asteroids. We chose this because the author has prior experience in explaining and demonstrating asteroid formation to student visitors of a public science center (CSSC). This made anecdotal assessment easier. All results of the methods mentioned below can be found in \href{https://github.com/yooleee/chatbot-research}{https://github.com/yooleee/chatbot-research}.}

\paragraph{The following strategies were explored to evaluate the research question.}

\begin{enumerate}
    \item Use of ChatGPT-3.5 to establish a baseline for Bloom’s taxonomy (BT) – i.e. how much does ChatGPT know about BT without additional training.
    \item Use of OpenAI Playground and \textit{GPT-3.5-turbo-instruct} to evaluate QG using BT
    \item Use of \textit{GPT-3.5-turbo-instruct} to generate persona-based chats
    \item Use of \textit{GPT-3.5-turbo-instruct} to combine personas and QG in a single chatbot session
    \item Use of \textit{GPT-3.5-turbo-instruc}t to evaluate the categories of BT questions as a means to check if the questions were coherent with the categories – i.e. using AI to check AI.
\end{enumerate}

\paragraph{The experiments were carried out regardless of any user interface (UI) design considerations. This poses a limitation upon the work because there are reasons to believe that different UI configurations might yield different results. This became clearer during one of the experiments wherein questions of the BT-Remembering category were generated prior to exposing the student to the materials contained in the questions.}

\subsection*{Bloom’s Taxonomy (BT)}

\paragraph{Our starting point was to be guided by the work of Elkins et al., (2024) in their paper that explored QG using BT. They used two methods: simple prompting strategy and controlled prompting strategy. The former merely presented the passage to the model and prompted it to generate 6 questions:}

\begin{verbatim}
Generate 6 questions.}
Passage: {context}
Questions:
\end{verbatim}

\paragraph{We noted that in previous work, the authors had explored the ideal length of these contexts, but we did not adhere to any such guidance. However, this would be useful to evaluate in future work. We also noted that they had not evaluated how the source text might influence the ability, or not, to support the generation of questions reliably for all six BT categories. Again, we would like to evaluate this in future work.}

\paragraph{In our work, we only paid attention initially to the more elaborate controlled prompting strategy as, according to the authors, this had proven more effective in generating questions. The controlled prompting strategy uses a technique known as few-shot learning (Parnami et al., 2022), which is also explained in OpenAI’s \href{https://platform.openai.com/docs/guides/prompt-engineering/strategy-give-models-time-to-think/}{Prompt Engineering Guide} and as shown in figure 1.}

\begin{figure}[h!tbp]
    \centering
    \includegraphics[width=1.0\linewidth]{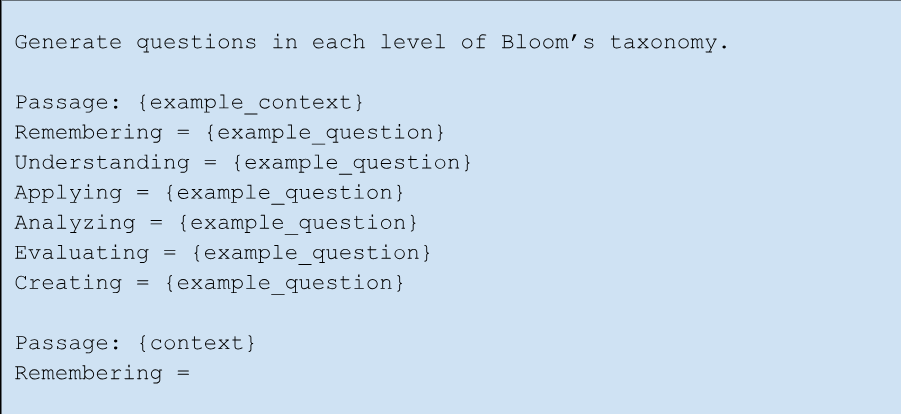}
    \caption{Example Template of Few-Shot Learning from Parnami et al.}
    \label{fig:1}
\end{figure}

\paragraph{However, after consulting with OpenAI’s \href{https://help.openai.com/en/articles/6654000-best-practices-for-prompt-engineering-with-openai-api}{Best Practices} for prompt engineering, we decided first to attempt zero-shot learning. In other words, how well does a state-of-the-art model already understand BT? For quick evaluation, we used ChatGPT-3.5 to complete the prompt “Please summarize Bloom's taxonomy and give examples of each type of question”. The results (see transcript) revealed that ChatGPT had existing knowledge of BT, although the answers given were typically for the revised version of BT (Anderson et al., 2001) (Wilson, 2016), as shown in figure 2.}

\paragraph{We ran experiments to compare zero-shot QG with Elkins’ controlled prompt strategy, using some of the original data (see \href{https://anonymous.4open.science/r/EQG_in_practice-2752/raw_data/few_shot_examples_ML.csv}{anonymous repo}) from Elkins. We noted that whilst the zero-shot was capable of producing credible questions from the sample text, sometimes the apparent category of generated questions strayed from the official taxonomic categories. That said, it is not always clear whether or not a particular question is fully contained within a single category or straddles category boundaries. Moreover, it is not clear in Elkins’ original research that the use of few-shot learning succeeds in all cases in providing a basis for which the GenAI can extrapolate to new examples. The evaluation method in Elkins was human QG versus GenAI QG, not a complete analysis of the efficacy of GenAI in generating questions that are always conformant with Bloom’s categories in each case.}

\begin{figure}[h!tbp]
    \centering
    \includegraphics[width=1\linewidth]{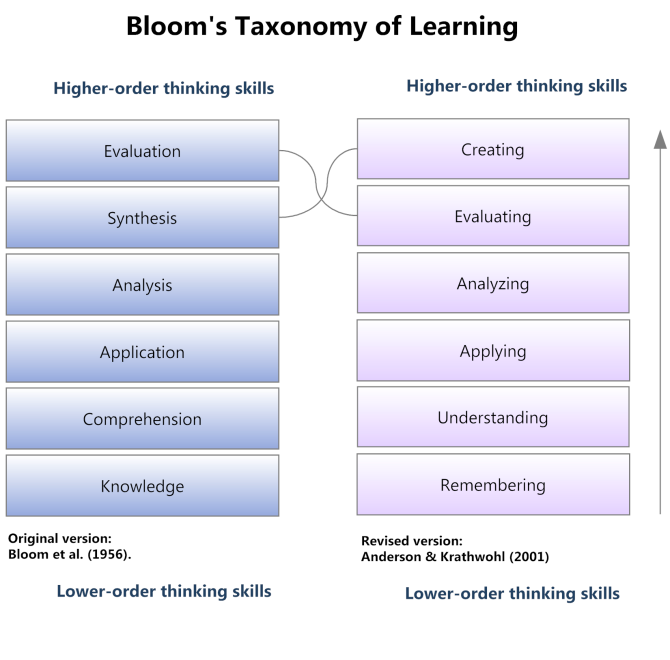}
    \caption{Bloom's Taxonomy: Original (left) and revised (right) -- \href{https://joanakompa.com/2017/02/07/why-it-is-time-to-retire-blooms-taxonomy/}{Source}}
    \label{fig:2}
\end{figure}

\paragraph{Our conclusion was that zero-shot provided sufficiently good performance to proceed with combining with the personified-chatbot approach, noting that the areas of difficulty in zero-shot were more often in the higher-order categories. This was a useful observation because it was decided that for an initial chatbot prototype, limiting the questions to only the first two levels would be more appropriate to a learning experience limited to a single chat as opposed to a more prolonged blended-learning experience that might warrant exploring the higher order categories.}

\paragraph{It is not clear why Elkins’ paper did not attempt zero-shot QG directly using BT within the prompt (versus the open-ended, or non-BT, simple prompting strategy) but we note that their experiments were conducted using \textit{text-davinci-003} which is an earlier LLM. Had they had access to more recent models, perhaps zero-shot would have been explored more fully.}

\subsection*{Science Lessons: Data Generation}

\paragraph{Having tested zero- versus few-shot learning and concluded to use zero-shot for the more extended research, we wanted to test zero-shot on the proposed domain of school-level science. Initially, we scraped content from the NASA educational guides (e.g. \href{https://www.jpl.nasa.gov/edu/teach/activity/modeling-an-asteroid/}{Modeling an Asteroid}). Note that we did this for experimental reasons only, not intended to commercialize the use of such materials. Hence we did not seek permission. Our final goal, if we were to develop a tool, would be to make it available to educationalists who presumably have access to their own materials, or could legally source them.}

\paragraph{However, it occurred to us to try generating the educational materials using ChatGPT given the following:}

\begin{enumerate}
    \item Large Language Models (LLMs) like GPT-3.5 have been proven to be so vast that they contain highly specialized knowledge bases (Veseli et al., 2023) across many subjects, even complex subjects like medicine (Nori et al., 2023). Hence, they surely must know about science (Cooper G., 2023).
    \item Models like GPT-3.5 have been demonstrated to be effective in generating materials aimed at certain levels of understanding – e.g. the famous prompt of ELI5 (Explain Like I am 5 (years old)) (Valentini, Maria, et al., 2023).
    \item If 1 and 2 are true, then this would align with the third point, which is that educationalists will often favor convenience in production of educational materials. If an LLM can do it, then why not incorporate such possibilities into the learning process, notwithstanding potential pitfalls like hallucinations (Bommasani et al., 2021).
    \item If materials can be adequately produced, per 1 and 2, whilst achieving convenience, per 3, then can they also be produced in a way that is already compatible with the need to generate related questions per BT?
\end{enumerate}

\paragraph{We found that the science-education materials produced by ChatGPT were seemingly adequate as a starting point for building a chatbot. We were able to combine all of the above four goals into a single prompting strategy, such as:}

\begin{verbatim}
Please provide me with a block of text to describe the astronomical
object called a comet. The level of text should be appropriate
for middle-school children and appropriate for a basic science
lesson. Please include sufficient levels of detail to derive some
questions in each level of Bloom's taxonomy.
\end{verbatim}

\paragraph{However, we also tried alternatives, one of which generated a context that we felt suitable for ongoing tests.\newline}

\begin{verbatim}
I am a volunteer at a space research visitor center that has many
school children visiting to learn about space and science. I want
to give some basic information for middle-schoolers about comets.
Please provide a basic information sheet.
\end{verbatim}

\paragraph{The \href{https://github.com/yooleee/chatbot-research/blob/main/prompts/comet_content_response.txt}{results}
 (which we call Subject Text) provided enough information for basic interaction and included some interesting elements about the origins of comets, their constituent parts and some of the more famous named examples (e.g. Halley’s Comet). Of course, in a more systematic and wider study within the context of classrooms, the production of the text and its suitability for powering a chatbot would need to be more formally evaluated. Note that the above prompt does not contain the BT prompting method, but we found a way of incorporating this later (see below).}

\subsection*{Anthropomorphic Design}

\paragraph{Recall, our goals were two-fold:}

\begin{enumerate}
    \item Get the chatbot to act in the first person about the object of learning – e.g., as if the user of the chatbot is chatting directly to a comet (as a person).
    \item Give the chatbot a particular voice, aligned with some persona.
\end{enumerate}

\paragraph{For evaluation, we used the OpenAI playground via the \href{https://platform.openai.com/docs/guides/text-generation/chat-completions-api}{Chat Completions API}. A prompt style that we discovered as useful is as follows:}

\begin{verbatim}
System:

You are a helpful assistant who knows about <subject> and can
teach young children who will ask questions. You will give
replies as if you are the comet, called <name>.

<persona hint>

What you know about <subject> is as follows:

<subject text>
\end{verbatim}

\paragraph{For the comet example, we populated the fields with the data from the table in Figure 3.}

\begin{figure}[htbp]
    \centering
    \includegraphics[width=1\linewidth]{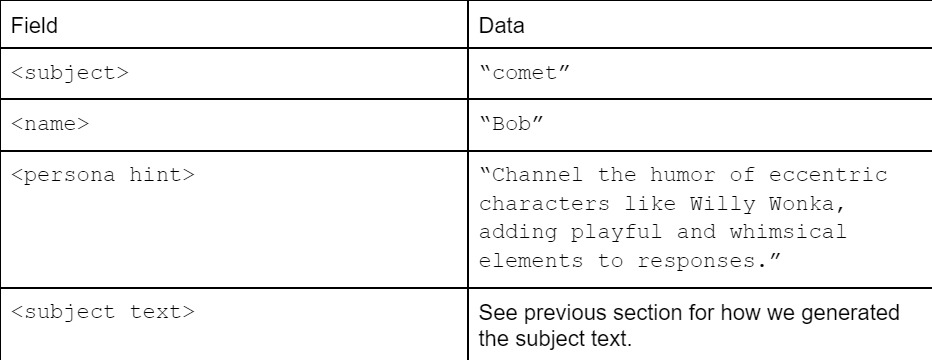}
    \caption{Data Used to Populate Prompt Placeholders}
    \label{fig:Table1}
\end{figure}

\paragraph{We noticed that in the chat exchanges, the persona was not always particularly noticeable. And we also noticed that the dialog was not in the first person as if the user were chatting to a comet. Example:}

\begin{verbatim}
No worries! Comets are composed of a mixture of ice, dust, and rock.
They are often referred to as "dirty snowballs" or "icy dirtballs"
because of their unique combination of materials. The ice in comets
can be frozen water, methane, ammonia, and other gases, while the
dust and rock make up the solid core known as the nucleus.
\end{verbatim}

\paragraph{The text "Comets are.." should have ideally said: “I am…”. We solved these problems using two fixes. Firstly, for the persona hint we made a subtle change:}

\paragraph{Before:}

\begin{verbatim}
"Channel the humor of eccentric characters like Willy Wonka,
adding playful and whimsical elements to responses.”
\end{verbatim}

\paragraph{Revision to emphasize first-person behavior:}

\begin{verbatim}
"Channel the humor of eccentric characters as if you are
Willy Wonka, adding playful and whimsical elements to responses.”
\end{verbatim}

\paragraph{To include more emphasis of the use of the persona’s voice throughout the chat session, we added a further modification to the prompt:}

\begin{verbatim}
"Be sure to rewrite your responses in the humor of Willy Wonka.”
\end{verbatim}

\paragraph{It added quite significant persona-based flourishes to the text – (see \href{https://platform.openai.com/playground/p/9zytyQlBOhcTsytXaUZWRgNX?mode=chat}{playground}). Example:}

\begin{verbatim}
"Welcome, young astronomers! I'm Bob, the lovable comet here to
dazzle you with my sparkling knowledge. So let's blast off into
the galaxy of information about comets!

Did you know that comets are like cosmic snowballs? Yes, indeed!
They have a magical mix of ice, dust, and rock. Just imagine a
snowball made of glitter and stardust. Ah, it's simply marvelous!

Now, my young friends, I have a question to tickle your brain.
Do you remember where comets come from? Are they freshly baked
in my cosmic confectionary? Or do they come from somewhere
far, far away?”
\end{verbatim}

\paragraph{Although we restricted our research purely to the generation of lower-ranking BT questions, there is research to indicate that LLMs can provide suitable feedback to open-ended questions (Matelsky et al., 2023). In the future, we would like to conduct further research into this to see how well the LLM can cope with higher-order open questions framed via Bloom’s Taxonomy.}

\subsubsection*{Hybrid Persona and Embedded Questions}

\paragraph{For this test, we attempted to incorporate some basic Qs (from BT) into the chat itself. For simplicity, we initially constrained Qs to only the first two categories: (\textit{understanding}, \textit{remembering}). What we found is that the bot inserted Qs after every interaction but without necessarily having introduced the relevant content from which the Q was derived.}

\paragraph{This leads to a research question: could we keep track of what content has been discussed in the chat session and confine our questions only to the historical data (not the general prompt content – \textit{subject text}). On the surface, this would seem possible if we use a programmatic method to call the API and only feed in student-seen data. However, without a wider set of data for the \textit{subject text} field, we don’t know how to handle, in advance, particular interests the student might have. In the long run, for a field-usable chatbot we would consider other methods to inject data dynamically into the prompt, but that was out of scope for the current research.}

\paragraph{However, there is a wider discussion about whether or not it makes sense to display general content about \textit{subject} as part of the UX–i.e. Outside of the chatbot session interface in a so-called “blended-learning” interface (see section below on Interface Design).}

\paragraph{For the questions generated, we noted that the bot was able to construct valid questions that were aligned with the taxonomy. For example, one question was:\\}

\begin{verbatim}
Now, here's a question for you: Which region in space is farther
away from our solar system, the Oort Cloud or the Kuiper Belt?
\end{verbatim}

\paragraph{This is a question of understanding, although perhaps one of analysis (bearing in mind our earlier remarks about the potential overlap between categories). But it shows how GenAI can use sufficient reasoning to pose understanding questions. Based upon the source text, this question is answerable, but requires the student to exercise understanding and reasoning by realizing that because Neptune is in the solar system, the Kuiper belt is therefore closer and so, per the line of questioning, the Oort cloud is further away. Note that the GPT model was able to answer this using reasoning as the answer for this question isn’t explicitly said in the source text. Here is the chatbot answer to the response Oort cloud:\newline}

\begin{verbatim}
Absolutely right! The Oort Cloud is indeed farther away from
our solar system compared to the Kuiper Belt. It's like the
ultimate cosmic hideout for comets, located at an extreme
distance from the Sun. The comets that originate from the
Oort Cloud can take a really long time to travel towards
the inner solar system, which adds to their sense of mystery
and intrigue.
\end{verbatim}

\paragraph{Note that most of our testing involved entering affirmative (correct) answers to questions, due to time limitations. Further research is required to see what happens if the student gives incorrect answers. We say this because it would be important to tell the student where mistakes were made and how to rectify them. This would entail the chatbot needing to give explicit reasoning instructions, which seems a different GenAI task than just performing the reasoning steps internally.}

\subsection*{Interface Design}

\paragraph{Note that our research has been purely within the confines of the OpenAI playground (and ChatGPT) in order to explore the feasibility of the hypotheses. Of course, we do not know what form an actual chatbot interface design should take. Whilst ChatGPT itself is a purely dialog-based user interface (UI), its goal is not to provide an educational experience per se, but rather fulfill a general-purpose informational query paradigm. Nominally, we might imagine components of a more blended approach, such as shown in figure 4:}

\paragraph{There is clearly work to be done in assessing viable chatbot UI designs (Haugeland, Isabel Kathleen Fornell, et al., 2022), which we believe could be researched using our work combined with Design Thinking (Micheli, Pietro, et al., 2019). As with any product design, some experimental basis is useful to inform an approach. Our experiments highlighted some potential UI insights:}

\begin{enumerate}
    \item The prompts that we explored sometimes resulted in questions being asked about content from the \textit{subject text} field that had not yet been revealed via the chat session. One possibility is to include the content as Support Materials (see UI diagram) such that it can be referenced outside of the chat box.
    \item  In some research (Anamalai et al., 2023)  it was found that student motivation was higher (within the motivational framework Self Determination Theory) if the chatbot was used more for assessment than learning. This might suggest the usefulness of tracking or presenting questions separately, per the “Generated Qs'' box in the nominal UI.
\end{enumerate}

\begin{figure}
    \centering
    \includegraphics[width=0.8\linewidth]{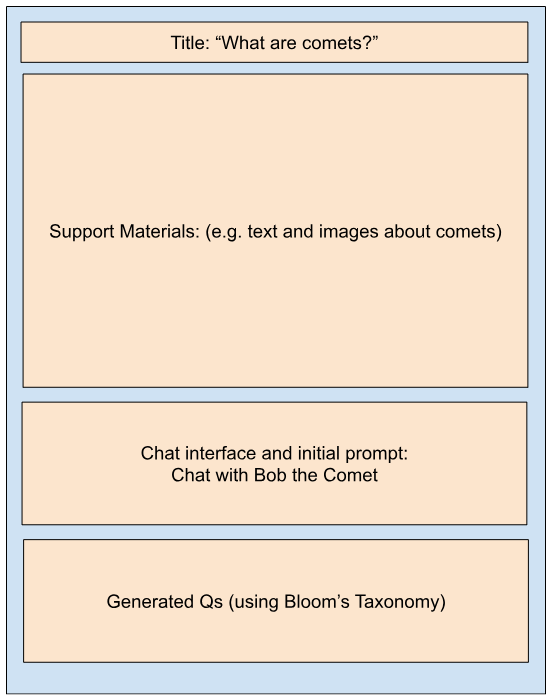}
    \caption{Potential User Interface Components (Conceptual Only)}
    \label{fig:Example_Overlay}
\end{figure}

\section*{Conversational Competence}

\paragraph{We did not have time to evaluate the conversational competency of the chatbots using the prompts provided. By conversational competency, we mean the ability of the chatbot to meaningfully sustain a conversation for the duration of the session, as dictated by the student, in whatever order the student wishes to take. This would be a critical area of research to understand how well the chatbot can maintain competence within the realms of what we are exploring, namely keeping the student engaged about the subject and via the benefits of incorporating the educational benefits of Bloom’s Taxonomy.}

\paragraph{Within the limited time and confines of this anecdotal study, we did explore what happens when the student either gets stuck (“I don’t know”) or provides incorrect answers. These examples are contained in the accompanying repo. We make the observation here that the ability for the chatbot (via OpenAI’s playground chat model using \textit{GPT-3.5-turbo-instruct}) to maintain a meaningful conversation without complete degradation was quite surprising. However, we did not push this to its limits and would plan to do so in future evaluations.}

\section*{Proposed Methods of Field Evaluation}

\paragraph{As noted, due to lack of resources (including time), our initial research question was limited to the exploration of the feasibility of incorporating anthropomorphism and Bloom’s Taxonomy into a chatbot interface using GenAI prompt engineering, as discussed in the preceding sections. However, we realize this is insufficient to draw any empirically useful pedagogical conclusions. So here we outline what we might do to extend the work and properly evaluate the proposed methods within the context of field-based educational circumstances. Using the different prompts, we would ask a group of students and teachers to interact with the bot and (with their permission) store transcripts of every chat session for subsequent analysis.}

\subsection*{General Evaluation}

\paragraph{Critical things we might need to evaluate include:}

\begin{enumerate}
    \item How often does the student appear to understand the Q and get it right?
    \item Are the generated Qs appropriate for the prompt – i.e. are they from the desired BT categories of Q (e.g. \textit{understanding, remembering})?
    \begin{itemize}
        \item Could we test this using another prompt that tries to classify questions according to the taxonomy?
    \end{itemize}
    \item Are the Qs answerable given the text?
    \begin{itemize}
        \item Human alignment needed to check this?
        \item Can another prompt do this?
    \end{itemize}
    \item Is the persona evident and is it aligned with the prompt?
    \begin{itemize}
        \item Does this text match this persona?
        \item Is the response aligned with the prompt?
    \end{itemize}
    \item Is the persona effective?
    \begin{itemize}
        \item We could survey the users and teachers - rating of the persona?
        \item We could use pairwise assessment to rank responses.
        \item How engaging is the session? (Use A/B testing – those with the persona and those without.)
        \begin{itemize}
            \item How long did they use it for? This seems more aligned to engagement – i.e. are they enjoying it?
            \item How many Qs did they get right? This seems more aligned to teaching – i.e. are they understanding it
        \end{itemize}
        \item What level of “persona intensity” improves engagement? (We could A/B test different levels of “persona intensity”.)
    \end{itemize}
    \item Learning – How well did they retain the materials based upon the use of the chatbot? Various tests could be formulated, as in giving children an interface with static content and Qs versus giving students the bot. We could test the students on the subject some time later to see if they have retained any info if we are focussed upon the (\textit{remembering}) aspects of Bloom’s taxonomy.
\end{enumerate}

\subsection*{Field Evaluation in the Context of Virtual Exhibits}

\paragraph{In addition to the above methods of evaluation, we would propose further evaluation criteria to explore the following:}

\begin{enumerate}
    \item How effective is the use of a chatbot as a “virtual exhibit” as part of a wider set of exhibits and educational experiences offered by a field-trip?
    \item How best could a virtual exhibit be combined with a physical field-trip?
    \item Can the use of the chatbot virtual exhibit provide adequate engagement with science in the absence of a field-trip (e.g. for cost-saving reasons)?
    \item Is it indeed the case that educationalists and volunteers within public-science institutions could easily produce their own virtual exhibits using the methods discovered in our research?
\end{enumerate}

\subsection*{Inclusivity Evaluation}

\paragraph{Given our opening remarks about the potential for field-trips and GenAI to enhance and extend STEM inclusion to include marginalized groups, further evaluation along these lines is needed:}

\begin{enumerate}
    \item Which personas are best suited to a particular cultural and community context to enhance the goals of STEM inclusion?
    \item More generally, how can the data-generation aspect (to produce Subject Text) be tailored to culturally sensitive and inclusive needs, leveraging ongoing work of groups like the \href{https://ceisl.iupui.edu/index.html?_gl=1*1ig9u2q*_ga*MTk2NjI4NDU4NS4xNzA2MTMyMTE3*_ga_61CH0D2DQW*MTcwNjEzODg0MC4zLjAuMTcwNjEzODg0MC42MC4wLjA.&_ga=2.120975304.1362142678.1706132117-1966284585.1706132117}{Collaborative for Equitable and Inclusive STEM Learning} (CEISL)?
    \item What is the impact of virtual exhibits upon marginalized groups who might otherwise be excluded from field-trips?
\end{enumerate}

\section*{Potential Improvements for Engagement}

\paragraph{Whilst we explored a relatively narrow research question of possible GenAI pathways to implementing virtual science exhibits using chatbots, we recognize there are many potential improvements directly related to the goal of increasing engagement. Our research showed the highly flexible nature of GenAI-based chatbots and the fluidity by which GenAI can produce contextually appropriate questions. However, whilst this flexibility suggests a potential benefit to students in allowing them to engage with the subject matter in any order, there is no evidence to suggest that this will keep a student motivated to keep learning. This needs careful research, per our preceding remarks about evaluation.}

\paragraph{We suggest that a possible improvement to the wider scheme is to consider the incorporation of other mechanisms designed to keep students motivated. One such mechanism is gamification. The work of Nuemann et al., (2023) suggests evidence of increased engagement via the tailored gamification in educational chatbots, particularly among users with gaming experience. With time, we would like to evaluate whether or not GenAI “understands” gamification via prompt engineering techniques. Of course, newer GenAI capabilities include the generation of images, 3D models and video elements. Therefore, it would be useful to explore the production of richer interfaces, like 3D avatars (Bai et al., 2024), for improved student engagement.}

\section*{Conclusion}

\paragraph{Our research demonstrates that using GenAI and prompt-engineering, scientific objects, such as comets, can be personified as first-person chatbots with customizable personalities. This is the first step towards creating engaging virtual science exhibits. We also found that the prompts can be extended to generate contextual relevant questions aligned with Bloom’s taxonomy. This paves the way to more engaging modes of learning via appropriate and educationally useful question generation. These findings could form the basis for making virtual exhibitions available to those alienated from science field-trips. Future research should concentrate on comprehensive in-field classroom evaluations, in conjunction with exhibition owners whilst exploring refinements to incorporate anthropomorphized styles suitable for marginalized groups. We believe that this research points to the possibilities for leveraging GenAI to foster accessible and inclusive approaches to public science education.}

\section*{Acknowledgements}
\paragraph{The author would like to thank the advice of Ahmer Mumtaz, CEO of Higher Summit. He provided invaluable guidance about the inclusion of Bloom's Taxonomy and the potential for first-person personification of exhibits.}

\section*{References}

\paragraph{Ait Baha, Tarek, et al. "The power of personalization: A systematic review of personality-adaptive chatbots." SN Computer Science 4.5 (2023): 661.}
\paragraph{Anderson, Lorin W., and David R. Krathwohl. A taxonomy for learning, teaching, and assessing: A revision of Bloom's taxonomy of educational objectives: complete edition. Addison Wesley Longman, Inc., 2001.}
\paragraph{Annamalai, Nagaletchimee, et al. "Exploring English language learning via Chabot: A case study from a self determination theory perspective." Computers and Education: Artificial Intelligence (2023): 100148.}
\paragraph{Bai, Song, and Jie Li. "Progress and Prospects in 3D Generative AI: A Technical Overview including 3D human." arXiv preprint arXiv:2401.02620 (2024).}
\paragraph{Baidoo-Anu, David, and Leticia Owusu Ansah. "Education in the era of generative artificial intelligence (AI): Understanding the potential benefits of ChatGPT in promoting teaching and learning." Journal of AI 7.1 (2023): 52-62.}
\paragraph{Behrendt, Marc, and Teresa Franklin. "A review of research on school field trips and their value in education." International Journal of Environmental and Science Education 9.3 (2014): 235-245. }
\paragraph{Bommasani, Rishi, et al. "On the opportunities and risks of foundation models." arXiv preprint arXiv:2108.07258 (2021).}
\paragraph{Cooper, Grant. "Examining science education in chatgpt: An exploratory study of generative artificial intelligence." Journal of Science Education and Technology 32.3 (2023): 444-452. }
\paragraph{Elkins, Sabina, et al. "How Teachers Can Use Large Language Models and Bloom's Taxonomy to Create Educational Quizzes." arXiv preprint arXiv:2401.05914 (2024)}
\paragraph{Ekwueme, Cecilia O., Esther E. Ekon, and Dorothy C. Ezenwa-Nebife. "The Impact of Hands-On-Approach on Student Academic Performance in Basic Science and Mathematics." Higher education studies 5.6 (2015): 47-51.}
\paragraph{Gozalo-Brizuela, Roberto, and Eduardo C. Garrido-Merchán. "A survey of Generative AI Applications." arXiv preprint arXiv:2306.02781 (2023).}
\paragraph{Janson, Andreas. "How to leverage anthropomorphism for chatbot service interfaces: The interplay of communication style and personification." Computers in Human Behavior 149 (2023): 107954.}
\paragraph{Kompa, Joana Stella. "Why it is time to retire Bloom’s taxonomy." Retrieved December 5 (2017): 2020.}
\paragraph{Matelsky, Jordan K., et al. "A large language model-assisted education tool to provide feedback on open-ended responses." arXiv preprint arXiv:2308.02439 (2023).}
\paragraph{Neumann, Alexander Tobias, et al. "Motivating Learners with Gamified Chatbot-Assisted Learning Activities." International Conference on Web-Based Learning. Singapore: Springer Nature Singapore, 2023.}
\paragraph{Nori, Harsha, et al. "Can generalist foundation models outcompete special-purpose tuning? case study in medicine." arXiv preprint arXiv:2311.16452 (2023)}
\paragraph{Pérez, José Quiroga, Thanasis Daradoumis, and Joan Manuel Marquès Puig. "Rediscovering the use of chatbots in education: A systematic literature review." Computer Applications in Engineering Education 28.6 (2020): 1549-1565.}
\paragraph{Parnami, Archit, and Minwoo Lee. "Learning from few examples: A summary of approaches to few-shot learning." arXiv preprint arXiv:2203.04291 (2022).}
\paragraph{Pizzi, Gabriele, Daniele Scarpi, and Eleonora Pantano. "Artificial intelligence and the new forms of interaction: Who has the control when interacting with a chatbot?." Journal of Business Research 129 (2021): 878-890.}
\paragraph{Sudirtha, I. Gede, I. Wayan Widiana, and Made Aryawan Adijaya. "The Effectiveness of Using Revised Bloom's Taxonomy-Oriented Learning Activities to Improve Students' Metacognitive Abilities." Journal of Education and e-Learning Research 9.2 (2022): 55-62.}
\paragraph{Tate, Tamara, et al. "Educational research and AI-generated writing: Confronting the coming tsunami." (2023).}
\paragraph{Valentini, Maria, et al. "On the Automatic Generation and Simplification of Children's Stories." arXiv preprint arXiv:2310.18502 (2023).}
\paragraph{Vázquez-Cano, Esteban, Santiago Mengual-Andrés, and Eloy López-Meneses. "Chatbot to improve learning punctuation in Spanish and to enhance open and flexible learning environments." International Journal of Educational Technology in Higher Education 18.1 (2021): 1-20.}
\paragraph{Veseli, Blerta, et al. "Evaluating the Knowledge Base Completion Potential of GPT." arXiv preprint arXiv:2310.14771 (2023)}
\paragraph{Whitesell, Emilyn Ruble. "A day at the museum: The impact of field trips on middle school science achievement." Journal of Research in Science Teaching 53.7 (2016): 1036-1054. }
\paragraph{Wilson, Leslie Owen. "Anderson and Krathwohl–Bloom’s taxonomy revised." Understanding the new version of Bloom's taxonomy (2016). }
\paragraph{Wu, Rong, and Zhonggen Yu. "Do AI chatbots improve students learning outcomes? Evidence from a meta‐analysis." British Journal of Educational Technology (2023).}
\paragraph{Winkler, Rainer, and Matthias Söllner. "Unleashing the potential of chatbots in education: A state-of-the-art analysis." Academy of Management Proceedings. Vol. 2018. No. 1. Briarcliff Manor, NY 10510: Academy of Management, 2018.}
\paragraph{Yin, Jiaqi, et al. "Conversation technology with micro-learning: The impact of chatbot-based learning on students’ learning motivation and performance." Journal of Educational Computing Research 59.1 (2021): 154-177.}

\end{document}